\documentclass[twocolumn]{article}

\usepackage[left=1in,right=1in,top=1in,bottom=1in]{geometry}
\usepackage{tikz}
\usetikzlibrary{calc,patterns,angles,quotes}

\usepackage{algorithm}
\usepackage{algpseudocode}
\usepackage{amsmath}
\usepackage{authblk}
\usepackage{url}
\usepackage{derivative}
\usepackage{booktabs}
\usepackage{graphicx}
\usepackage{caption}
\usepackage{subcaption}
\usepackage{hyperref}
\usepackage{float}
\usepackage{tikz}
\usepackage{enumitem}
\usetikzlibrary{positioning, arrows.meta}

\usepackage[explicit]{titlesec}
\renewcommand{\thesection}{\Roman{section}}
\renewcommand{\thesubsection}{\thesection.\Alph{subsection}}
\renewcommand{\thesubsubsection}{\thesubsection.\arabic{subsubsection}}

\titleformat{\section}{\large\scshape\centering}{\thesection.\space}{0pt}{#1}[]
\titlespacing*{\section}{0pt}{0.5\baselineskip}{0pt}
\titleformat{\subsection}{\normalsize\itshape}{\Alph{subsection}.\space}{0pt}{#1}[]
\titlespacing*{\subsection}{0pt}{0.5\baselineskip}{0pt}
\titleformat{\subsubsection}{\normalsize\itshape}{\arabic{subsubsection}.\space}{0pt}{#1}[]
\titlespacing*{\subsubsection}{0pt}{0.5\baselineskip}{0pt}
\usepackage{geometry}
\newgeometry{left=2cm, right=2cm, top=2.5cm,bottom=3.5cm}
\makeatletter
\renewcommand{\fnum@figure}{Fig. \thefigure}
\renewcommand{\fnum@table}{Tab. \thetable}
\makeatother

\usepackage{nopageno}
\renewcommand{\abstractname}{}
\title{\textbf{\normalsize Computing low-thrust transfers in the asteroid belt, a comparison between astrodynamical manipulations and a machine learning approach}}
\author[(1), (2)]{Giacomo Acciarini}
\author[(3)]{Laurent Beauregard}
\author[(1)]{Dario Izzo}
\affil[(1)]{Advanced Concepts Team, European Space Agency, European Space Research and TechnologyCentre (ESTEC), Keplerlaan 1, 2201 AZ Noordwijk, The Netherlands}
\affil[(2)]{Surrey Space Centre, University of Surrey, GU2 7XH, Guildford, United Kingdom}
\affil[(3)]{European Space Operations Centre, Darmstadt, 64293, Germany}
\affil[]{g.acciarini@surrey.ac.uk, dario.izzo@esa.int}
\date{}  

\begin{document}
\maketitle

\begin{abstract}
\vspace{-1.3\baselineskip}
\textbf{\emph{\quad Abstract} - 
Low-thrust trajectories play a crucial role in optimizing scientific output and cost efficiency in asteroid belt missions. Unlike high-thrust transfers, low-thrust trajectories require solving complex optimal control problems. This complexity grows exponentially with the number of asteroids visited due to orbital mechanics intricacies. In the literature, methods for approximating low-thrust transfers without full optimization have been proposed, including analytical and machine learning techniques. In this work, we propose new analytical approximations and compare their accuracy and performance to machine learning methods. While analytical approximations leverage orbit theory to estimate trajectory costs, machine learning employs a more black-box approach, utilizing neural networks to predict optimal transfers based on various attributes. We build a dataset of about 3 million transfers, found by solving the time and fuel optimal control problems, for different time of flights, which we also release open-source. Comparison between the two methods on this database reveals the superiority of machine learning, especially for longer transfers. Despite challenges such as multi revolution transfers, both approaches maintain accuracy within a few percent in the final mass errors, on a database of trajectories involving numerous asteroids. This work contributes to the efficient exploration of mission opportunities in the asteroid belt, providing insights into the strengths and limitations of different approximation strategies.}
\end{abstract}

\section{Introduction}
The number of asteroids in the asteroid belt larger than 1 km in diameter is estimated to be between 1 and 2 million~\footnote{\url{https://science.nasa.gov/solar-system/asteroids/facts/}, date of access: February 2024.}, making it the area of our Solar System with the majority of asteroids~\cite{demeo2015compositional}.
Studying these asteroids is crucial for several reasons, ranging from compositions and Solar System origin studies, to resource exploration, planetary defense, and impact and collision dynamics investigations~\cite{minton2009record, bottke2015collisional, farinella1992collision}. 
Among all space missions, only NASA's Dawn spacecraft~\footnote{\url{https://science.nasa.gov/mission/dawn/}, date of access: February 2024.} has explored celestial bodies within the main asteroid belt (i.e., Vesta and Ceres) over an extended duration~\cite{russell2012dawn},  while all other missions have only been brief flybys. Ideally, one would like to devise a single mission, to visit multiple asteroids of the same population: this has given rise to multiple flyby mission concepts. Most of the previous works that have studied this problem have proposed a mixed optimization procedure, with an inner continuous part (where transfers between asteroids are optimized for optimal time or mass) and an outer combinatorial part (where the best asteroid sequence is sought)~\cite{chen2014accessibility, grigoriev2013choosing, olympio2011optimal}. As shown by the Global Trajectory Optimization Competition (GTOC) series, which has featured several competitions involving multiple asteroid visits (including the very latest one~\footnote{\url{https://gtoc12.tsinghua.edu.cn/about}, date of access: February 2024.}), designing these missions can be very challenging, especially when the spacecraft is equipped with a low-thrust engine~\cite{izzo2016designing,vinko2008global, shen2023dyson}. Given this interplay between the outer combinatorial and inner continuous part, having methods to substantially decrease the burden of the inner optimization, while providing good initial guesses, becomes essential. Hence, researchers have been working on several techniques to approximate multi-impulsive and low-thrust transfers between asteroids since the seminal work of Edelbaum for time-optimal transfers~\cite{edelbaum1961propulsion, casalino2007improved}. Among these attempts, some examples also include the analytical efforts to find the maximum initial mass approximation (\textsc{MIMA}) for a low-thrust transfer, as well as several machine learning approaches~\cite{li2020deep}. Both analytical and machine learning models have the important advantage of substantially reducing the computational burden but might be limited in their generalization capability.

The objective of this work is twofold: first of all, we introduce new analytical approximations (building on the \textsc{MIMA}) and machine learning approximations. Then, we compare these techniques on a very large database of fuel and time optimal trajectories, to investigate the advantages and drawbacks of the two techniques. For this purpose, in Sec.~\ref{sec:methods}, the optimal control problems and dataset generation strategy, that have been used to generate the target data are discussed. Then, in Sec.~\ref{sec:analytical_approximations}
 and \ref{sec:machine_learning_approximations}, the new analytical and machine learning approximators are introduced, and finally, in Sec.~\ref{sec:numerical_experiments}, their accuracy on a database of roughly 3 million transfers is discussed.
\section{Methods}
\label{sec:methods}
\subsection{Optimal Control Problems}
Before developing any approximation, it is necessary to formulate and solve the corresponding optimal control problem (OCP): this solution will serve as the ground truth against which analytical and machine learning methods can be compared. We are interested in both time and fuel-optimal solutions. The time optimal solution will correspond to a trajectory without coast arcs, where the spacecraft will continuously thrust between the starting and arrival asteroids~\cite{pontryagin1962maximum}. This problem can be formulated as:
\begin{equation} 
\mathcal P_{tof^*}:
\left\{
    \begin{aligned}
            \mbox{given:} & \quad \textrm{ast}_s, \textrm{ast}_t, t_s \\
            \mbox{find:} & \quad \mathbf u(t) \\
        \mbox{to minimize:} & \quad t_t
        \\
        \mbox{subject to:} &\quad
        \mathbf{x}(t_s)-\mathbf{x}_{\textrm{ast}_s}(t_s)=\mathbf{0}\\
        &\quad \mathbf{x}(t_t)-\mathbf{x}_{\textrm{ast}_t}(t_t)=\mathbf{0}
        \\
        &\quad \dot{\mathbf y} = \mathbf f(\mathbf y, \mathbf u)\\
        &\quad |\mathbf u| \le 1\\
        &\quad m(t_s)=m_0
    \text{,}
    \end{aligned}
    \right.
    \label{eq:optimal_control_problem_min_tof}
\end{equation}
where $\pmb{x}$ is the vector of position and velocity in Cartesian coordinates (i.e., the spacecraft state), while $\pmb{y}=[\pmb{x}^T,m]^T$ represents the augmented state, while $t$ represents the epoch, $m_0$ the initial spacecraft mass, and $s$, $t$ are the subscripts that refer to the starting and target asteroids, respectively.
Basically, we choose starting and target asteroids, with fixed starting and target epochs: then, we solve for the minimum time of flight, for a given initial mass. This returns the minimum allowed time for transfer to be possible.
Note that for the solution of the time optimal problem of Eq.~\eqref{eq:optimal_control_problem_min_tof}, the mass of the spacecraft cannot be higher than the starting mass, otherwise the transfer would not be possible anymore with that time of flight: we will refer to this mass as maximum initial mass (MIM). In other words, for a fixed time transfer, that is the maximum mass allowed for the spacecraft, before the low-thrust transfer starts to become unfeasible.
Similarly, another OCP of interest is the fuel-optimal one. This consists of solving the OCP for a fixed time of flight, and optimizing for the consumed fuel:
\begin{equation} 
\mathcal P_{\Delta m^*}:
\left\{
    \begin{aligned}
            \mbox{given:} & \quad \textrm{ast}_s, \textrm{ast}_t, t_s, t_t \\
            \mbox{find:} & \quad \mathbf u(t) \\
        \mbox{to minimize:} & \quad J = \int_{t_s}^{t_t}u(t)dt 
        \\
        \mbox{subject to:} &\quad
        \mathbf{x}(t_s)-\mathbf{x}_{\textrm{ast}_s}(t_s)=\mathbf{0}\\
        &\quad \mathbf{x}(t_t)-\mathbf{x}_{\textrm{ast}_t}(t_t)=\mathbf{0}
        \\
        &\quad \dot{\mathbf y} = \mathbf f(\mathbf y, \mathbf u)\\
        &\quad |\mathbf u| \le 1\\
        &\quad m(t_s)=m_0
    \text{,}
    \end{aligned}
    \right.
    \label{eq:optimal_control_problem_min_fuel}
\end{equation}
The total $\Delta V$ and consumed fuel mass are connected to the final and initial mass (i.e., the consumed fuel) via Tsiolkovsky equation: $m_f/m_0=\textrm{exp}(-\Delta V / I_{sp}/g_0)$, where $g_0$ is the gravitational change at sea level (i.e., $g_0=9.80665 \ m/s^2$).

Both optimal control problems representing the low thrust trajectory are solved with a limited number of thrust arcs, following Sims approach~\cite{sims1997preliminary}, in conjunction with the utilization of the Non-Linear Programming solver SNOPT \cite{gill2005snopt, biscani2020parallel}. Moreover, twenty restarts for each OCP solution are used, to alleviate the problem of getting stuck in local optima.

\subsection{Dataset Generation}
\label{sec:dataset_generation}
Previous machine learning attempts have focused on transfers with some limitations, ranging from short time of flights to low eccentricities or low inclinations~\cite{mereta2017machine, izzo2019machine, li2020deep, viavattene2022artificial}. To comprehensively assess the performance of both analytical and machine learning methods, we have decided to devise the database more generically.
The baseline data was extracted from the database of the 12th GTOC competition: this comprises a set of 60,000 asteroids, in the region of the main asteroid belt. It is assumed that the spacecraft is equipped with an electric propulsion thruster, with a specific impulse of 4,000s, and a maximum thrust magnitude of 0.6 N. Furthermore, the initial mass of the spacecraft varied between 700 to 8,000 kg, with a resolution of 200 kg, to account for very different transfer geometries depending on how heavy the spacecraft is. Then, the time optimal control problem of Eq.~\eqref{eq:optimal_control_problem_min_tof}, was solved for about 1,7 million transfers: these were selected by randomly selecting the initial starting asteroid index and solving the time optimal control problem between that and the first five best neighbors, according to the orbital metric~\cite{hennes2016fast}.

Once this was done, for a random selection of these transfers, the equivalent fuel optimal control problem of Eq.~\eqref{eq:optimal_control_problem_min_fuel} was solved, by varying the time of flight in the interval $[1.2,1.78,2.35,2.93,3.5]tof^*$, where $tof^*$ is the optimal time of flight previously found. This amounted to about 1,3 million total transfers,
Both databases have been released open-source~\cite{acciarini_2024_10972838}.

Hence, a total of about 3 million transfers were created. This dataset was further divided into four different databases: $DB_{1}$ a low time of flight database of roughly 900,000 transfers, with time of flights ranging from a minimum of about 20 days, up to 365 days. $DB_2$ a moderate-low time of flight database of about 900,000 transfers, with time of flights ranging from a minimum of 365 days up to 700 days. $DB_3$ a moderate-high time of flight database of about 900,000 transfers, with time of flights ranging from 700 to 1,500 days. $DB_4$ a high time of flight database of about 300,000 transfers, with time of flights above 1,500 days.
The training and comparison of ML techniques against analytical approximations was performed on each database separately, as well as on the entire database (that comprises the four databases merged together).

\section{Analytical Approximations}
\label{sec:analytical_approximations}
As a first step, we approached the problem via analytical approximations: these are devised building on similar strategies as the maximum initial mass approximation (\textsc{MIMA})~\cite{hennes2016fast}. In particular, as already introduced in \cite{izzo_gtoc12_2024}, the idea was to enhance the above approximation with a more sensible estimate for the $\Delta V$, informed by Lambert's solution. 
Let's start by writing the augmented equations of motion of the spacecraft in terms of its Cartesian position and velocity:
\begin{align}
\begin{split}
\dot{\pmb{r}}&=\pmb{v}\\
\dot{\pmb{v}}&=\pmb{g}(\pmb{r})+T_{max}\dfrac{\pmb{u}(t)}{m(t)}\\
\dot{m}&=-T_{max}\dfrac{|\pmb{u}(t)|}{I_{sp}g_0}
\text{,}
\end{split}
\end{align}
where $|\pmb{u}(t)|\le1$, and $\pmb{g}(\pmb{r})$ is the gravitational acceleration exerted by the central body on the spacecraft. Then, we assume that the low-thrust trajectory can be treated as a perturbation around the Keplerian trajectory, hence we can write the equations that regulate deviations of the state as:
\begin{equation}
\delta \dot{\pmb{x}}=\dfrac{\partial \pmb{f}}{\partial \pmb{x}}\delta \pmb{x}+\begin{bmatrix}{\pmb{0}}\\T_{max}\dfrac{\pmb{u}(t)}{m(t)}\end{bmatrix}
\text{,}
\label{eq:delta_dot_x}
\end{equation}
where  $\pmb{x}=[\pmb{r}^T,\pmb{v}^T]^T$ is the state of the spacecraft and $\pmb{f}=\dot{\pmb{x}}$ is its dynamics. The solution of the homogeneous system of Eq.~\eqref{eq:delta_dot_x} is given by: \begin{equation}
    \dot{M}=\dfrac{\partial \pmb{f}}{\partial \pmb{x}}M
\end{equation}
where $M=\dfrac{\partial\pmb{x}}{\partial\pmb{x}_0}$ is the state transition matrix, which only depends on the reference trajectory, and in the case of Keplerian motion can be computed explicitly. Future deviations of the state (in the absence of thrust) can therefore be written as $\delta \pmb{x}=M\delta\pmb{x}_0$. By then applying "variation of parameters", we can differentiate this expression and combine it with Eq.~\eqref{eq:delta_dot_x} to get:
\begin{equation}
    \dot{M}\delta \pmb{x}_0+M\delta\dot{\pmb{x}}_0=\dfrac{\partial \pmb{f}}{\partial \pmb{x}}\delta \pmb{x}+\begin{bmatrix}{\pmb{0}}\\T_{max}\dfrac{\pmb{u}(t)}{m(t)}\end{bmatrix}
    \text{.}
\end{equation}
So, simplifying and solving for $\delta\pmb{x}_0$:
\begin{equation}
    \delta\dot{\pmb{x}}_0=M^{-1}\begin{bmatrix}{\pmb{0}}\\T_{max}\dfrac{\pmb{u}(t)}{m(t)}\end{bmatrix}
    \label{eq:eom_for_dx0}
\end{equation}
By knowing the thrust and mass profile, Eq.~\eqref{eq:eom_for_dx0} can be solved and future deviations of the state w.r.t. initial deviations can be mapped via the state transition matrix:
\begin{align}
\begin{split}
    \delta \pmb{x}(t)=&M(t)\delta \pmb{x}_0(t)=\\
    =&M(t)\int_{0}^tM^{-1}(s)\begin{bmatrix}{\pmb{0}}\\T_{max}\dfrac{\pmb{u}(s)}{m(s)}\end{bmatrix}ds+M(t)\pmb{C}
    \label{eq:delta_x_expression}
\text{,}
\end{split}
\end{align}
where $\pmb{C}$ is an integration constant.
As for the \textsc{MIMA} case, two assumptions are then made: first, that trajectory is composed of two thrust arcs, of duration $t_1$ and $t_2$ (while the total time of flight is indicated as $tof$), respectively. Secondly, the direction of the thrust vector is inertially fixed along each arc. It then follows that Eq.~\eqref{eq:delta_x_expression} can be further developed as:
\begin{align}
\begin{split}
        \delta \pmb{x}=M(t)&\bigg\{ \int_{0}^{t_1}M^{-1}(s)\begin{bmatrix}{\pmb{0}}\\T_{max}\dfrac{\pmb{u}_1(s)}{m_1(s)}\end{bmatrix}ds+\\
    &+\int_{tof-t_2}^{t_2}M^{-1}(s)\begin{bmatrix}{\pmb{0}}\\T_{max}\dfrac{\pmb{u}_2(s)}{m_2(s)}\end{bmatrix}ds+  \\
    &+M(t)\pmb{C}\bigg\}
\text{,}
\end{split}
\label{eq:expanded_dx}
\end{align}
where we also assume that the thrust magnitude is the same in the two thrust arcs (as for the MIMA case), and is always at its maximum (i.e., $|\pmb{u}_1(t)|=|\pmb{u}_2(t)|=T_{max}$):
\begin{align}
    \begin{split}
        m_1(t)&=m_0-\dfrac{T_{max}}{I_{sp}g_0}t\\
        m_2(t)&=m_0-\dfrac{T_{max}}{I_{sp}g_0}(t_1+t_2+t-tof)
\text{.}
\end{split}
\end{align}
We then impose the boundary conditions by ensuring that the final and initial position deviations w.r.t. the arrival and departing asteroids are zero and that their velocity gap corresponds to Lambert's solution: $\delta \pmb{x}_0=[\pmb{0},-\Delta\pmb{v}_{L,1}]^T$ and $\delta \pmb{x}_{tof}=[\pmb{0},\Delta\pmb{v}_{L,2}]^T$. By substituting these into Eq.~\eqref{eq:expanded_dx} and recalling that the thrust vectors are inertially fixed (i.e., $\pmb{u}_1(t)=\pmb{u}_1$,  $\pmb{u}_2(t)=\pmb{u}_2$) one obtains a system of six equations, which has the following form:
\begin{align}
    \begin{split}
&M^{-1}(tof)\delta \pmb{x}_{tof}-\delta \pmb{x}_0=\\
&=M^{-1}(tof)\begin{bmatrix}
        &\pmb{0}\\
&\Delta\pmb{v}_{L,1}
    \end{bmatrix}+\begin{bmatrix}
        \pmb{0}\\
        \Delta \pmb{v}_{L,2}
    \end{bmatrix}=\\
&=\int_{0}^{t_1}M^{-1}(s)\begin{bmatrix}
        \pmb{0}\\
        T_{max}\dfrac{\pmb{u}_1}{m_1(s)}
    \end{bmatrix}ds+\\
&+\int_{tof-t_2}^{tof}M^{-1}(s)\begin{bmatrix}
        \pmb{0}\\
        T_{max}\dfrac{\pmb{u}_2}{m_2(s)}
    \end{bmatrix}ds=\\
&=K_1 t_1 \pmb{u}_1+K_2t_2\pmb{u}_2
    \text{,}
    \end{split}
    \label{eq:dv_approx}
\end{align}
where $K_1, K_2$ are two matrices that will depend on $t_1$ and $t_2$, which can be computed by integrating the Keplerian state transition matrix multiplied by the mass: in our case, we use Simpson's method to approximate the integral. Hence, the only unknowns are the three components of the thrust vector direction as well as the two times. Then, the extra constraint that the thrust magnitude is maximum at each thrust arc can be added. These extra conditions allow us to have a system of eight equations in eight unknowns,  which can then be solved, to obtain $\pmb{u}_1$ and $\pmb{u}_2$, as well as $t_1$ and $t_2$.
Once this is done, it is finally possible to compute the cost of the approximated transfer arc, via:
\begin{equation}
    \Delta V\approx-I_{sp}g_0\textrm{ln}\bigg(1-\dfrac{T_{max} t_1+T_{max}t_2}{m_0I_{sp}g_0}\bigg)
    \text{,}
\end{equation}

A similar reasoning can also be used to have an approximation of the maximum initial mass. This can be done by leveraging the fact that in an optimal time transfer, there is no coast arc, hence: $tof=t_1+t_2$. In this case, the initial mass is assumed to be a variable, and solved for. This results in an approximation for the maximum initial mass, which we termed \textsc{MIMA2}~\cite{izzo_gtoc12_2024} since it has been developed as an enhancement of \textsc{MIMA}.
\section{Machine Learning Approximations}
\label{sec:machine_learning_approximations}
During the preliminary design of multiple rendezvous missions (including the GTOC series problems), many fuel and time-optimal OCPs similar to Eq.~\eqref{eq:optimal_control_problem_min_fuel} and \eqref{eq:optimal_control_problem_min_tof} are solved. This enables the possibility of using such data for training a machine learning model. Hence, our objective was to develop a lightweight model that can be trained on a database of transfers to predict their cost and to benchmark such models against the analytical approximations developed in Sec.~\ref{sec:analytical_approximations}. For this purpose, we used a feed-forward neural network architecture. Furthermore, to have a fair comparison between the two methods, we developed a strategy based on similar reasoning, and only used the transfer characteristics (i.e., the initial and final state of the spacecraft, initial mass, and time of flight) and the Lambert cost as attributes. In particular, the following attributes were used as inputs to the neural network:
\begin{enumerate}[itemsep=3pt,parsep=3pt,topsep=-3pt,partopsep=3pt]
    \item time of flight z-score normalized (i.e., subtracting the mean and dividing by the standard deviation);
    \item initial mass z-score normalized;
    \item relative position and velocity between initial and final point is computed using spherical coordinates, then the position and polar angles are z-score normalized, while the azimuth angle is fed in the network using sine/cosine (to account for the fact that 0 and 360 degrees azimuth angle correspond to the same points);
    \item initial and arrival Lambert costs (i.e., $\Delta \pmb{v}_{L,1}$, $\Delta \pmb{v}_{L,2}$), expressed in spherical coordinates and z-score normalized for the norm, and sine/cosine for the two angles;
    \item final and initial mass ratio using Lambert total $\Delta v$ as cost, and difference in mean motion and eccentricity, all z-score normalized. These are extra redundant attributes, which could be derived from the above inputs, but were nevertheless useful to maintain a rather small network architecture and have effective training.
\end{enumerate}
These resulted in a total of 23 inputs to the network.
The output of the network was the final to initial mass ratio: hence, the final activation was chosen to be the sigmoid, which forces the output to always be in the zero to one bounds. The schematic illustration of our small architecture is displayed in Fig.~\ref{fig:network_schematic_illustration}: three hidden layers and 32 neurons per layer were chosen. The focus was not on optimizing all hyperparameters of the network (such as the depth and length of the hidden layers, activation functions, etc.). Instead, our focus was on maintaining a small network size, having fast inference times, and being able to leverage these approximators during the combinatorial part of multiple asteroid tour design. Apart from the sigmoid in the output layer, ReLU was otherwise used as an activation function, and the training was performed with always the same network size (for all four databases, as well as for the overall dataset made of the union of the four). This resulted in a total amount of 2,913 learnable parameters (including weights and biases). The inference speed of the network on a single datapoint, including normalization and preparation of the inputs, is roughly 50$\times$ faster than the analytical approach discussed in Sec.~\ref{sec:analytical_approximations}.

\tikzset{
    neuron/.style={circle,draw,minimum size=1cm,inner sep=0pt},
    input/.style={neuron, fill=blue!20},
    hidden/.style={neuron, fill=green!20},
    output/.style={neuron, fill=red!20},
    conn/.style={-Stealth, thick},
}

\begin{figure}
\centering
\begin{tikzpicture}[scale=0.75, every node/.style={transform shape}]
    \foreach \i/\text in {1/{$\pmb{x}_t-\pmb{x}_s$}, 2/{$\Delta \pmb{v}_{L,1}$}, 3/{$\Delta \pmb{v}_{L,2}$}, 4/{$m_0$}, 5/{$tof$}} {
        \node[input] (I\i) at (0,\i*2.65) {\text};
    }
    
    \foreach \i in {1,...,2} {
        \node[hidden] (H1\i) at (2, \i*4) {};
    }
    \node at (3, 6) {$\vdots$};
    \foreach \i in {3,...,5} {
        \node[hidden] (H1\i) at (2, \i*2+2) {};
    }
    \draw[dotted, line width=1.5, gray!60!gray, rounded corners] (1.3, 12.8) rectangle (2.7, 3.2) node[pos=.5, yshift=5.2cm] {23$\times$32};
    
    \foreach \i in {1,...,2} {
        \node[hidden] (H2\i) at (4, \i*4) {};
    }
    \node at (6, 6) {$\vdots$};
    \foreach \i in {3,...,5} {
        \node[hidden] (H2\i) at (4, \i*2+2) {};
    }
    \draw[dotted, line width=1.5, gray!60!gray, rounded corners] (3.3, 12.8) rectangle (4.7, 3.2) node[pos=.5, yshift=5.2cm] {32$\times$32};

    \foreach \i in {1,...,2} {
        \node[hidden] (H3\i) at (6, \i*4) {};
    }
    \node at (6, 6) {$\vdots$};
    \foreach \i in {3,...,5} {
        \node[hidden] (H3\i) at (6, \i*2+2) {};
    }

    \draw[dotted, line width=1.5, gray!60!gray, rounded corners] (5.3, 12.8) rectangle (6.7, 3.2) node[pos=.5, yshift=5.2cm] {32$\times$1};
    
    \node[output] (O) at (8, 8) {$m_f/m_0$};
    
    \foreach \i in {1,...,5} {
        \foreach \j in {1,...,5} {
            \draw[conn] (I\i) -- (H1\j);
        }
    }
    
    \foreach \i in {1,...,5} {
        \foreach \j in {1,...,5} {
            \draw[conn] (H1\i) -- (H2\j);
        }
    }
    \foreach \i in {1,...,5} {
        \foreach \j in {1,...,5} {
            \draw[conn] (H2\i) -- (H3\j);
        }
    }
    
    \foreach \i in {1,...,5} {
        \draw[conn] (H3\i) -- (O);
    }
    
\end{tikzpicture}
\caption{Schematic illustration of neural network architecture.}
\label{fig:network_schematic_illustration}
\end{figure}
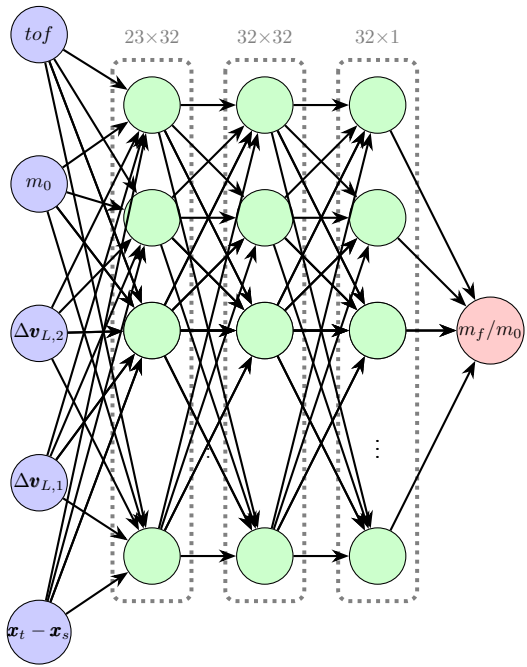
\section{Numerical Experiments}
\label{sec:numerical_experiments}
As mentioned in Sec.~\ref{sec:dataset_generation}, we tested both the analytical and machine learning techniques on four different databases, found by solving fuel and time optimal control problems in the asteroid belt and grouping the solutions by the time of flight. The machine learning approximator was trained for each database, and the results in terms of final mass estimation accuracy were compared with the analytical approximation. Furthermore, a comparison across the whole database (composed of the union of the four databases) was also made: in this case, the machine learning model was trained over all the data.

\begin{figure*}[hbt!]
  \centering
  \begin{subfigure}{0.49\textwidth}
    \includegraphics[width=\linewidth]{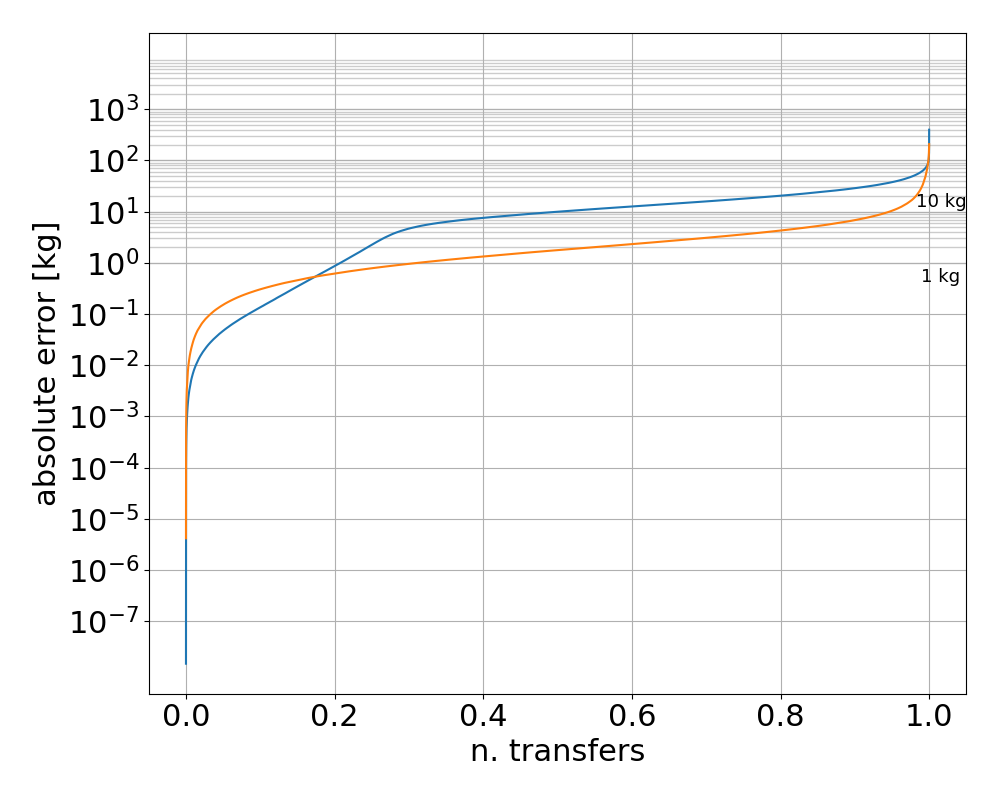}
    \caption{Low time of flight (between 0 and 365 days) database results.}
    \label{fig:abs_error_low_tof}
  \end{subfigure}
  \hfill
  \begin{subfigure}{0.49\textwidth}
    \includegraphics[width=\linewidth]{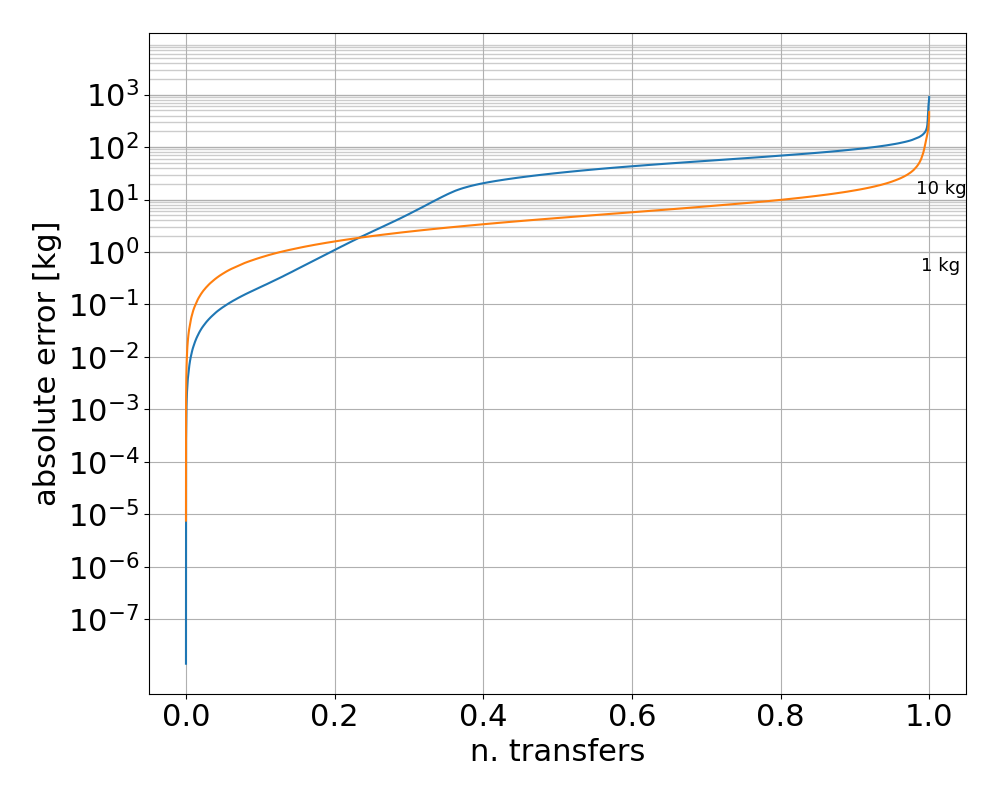}
    \caption{Medium-low time of flight (between 365 and 700 days) database results.}
    \label{fig:abs_error_medium_low_tof}
  \end{subfigure}
  \hfill
  \begin{subfigure}{0.49\textwidth}
    \includegraphics[width=\linewidth]{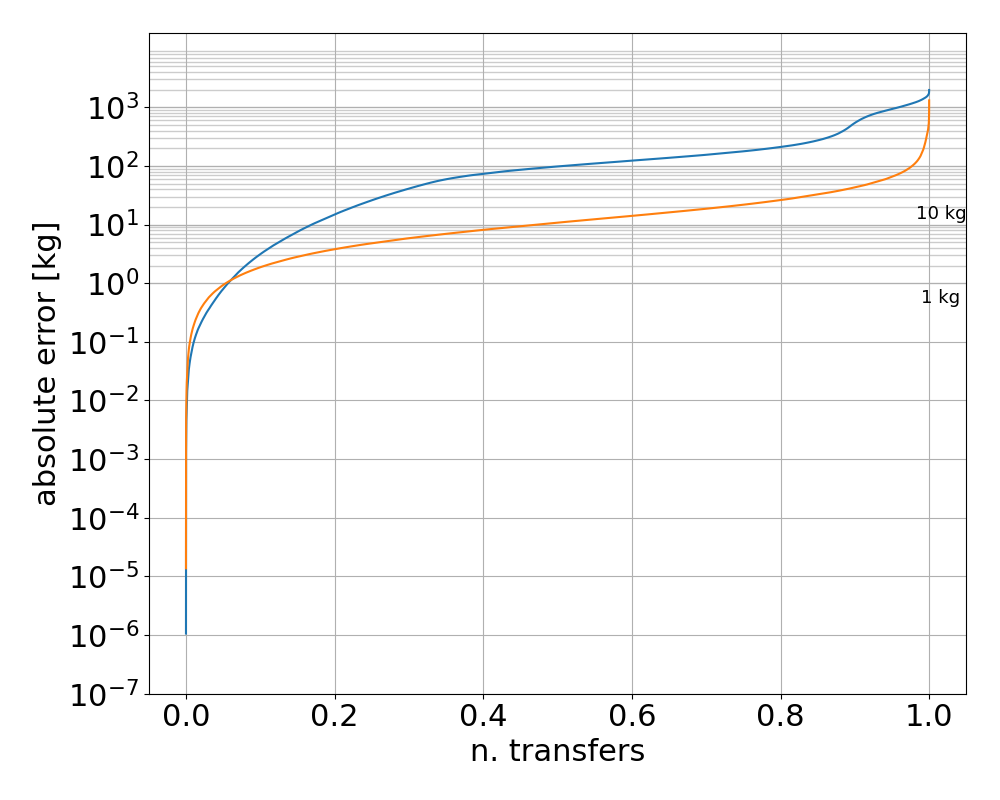}
    \caption{Medium-high time of flight (between 700 and 1500 days) database results.}
    \label{fig:abs_error_medium_high_tof}
  \end{subfigure}
  \begin{subfigure}{0.49\textwidth}
    \includegraphics[width=\linewidth]{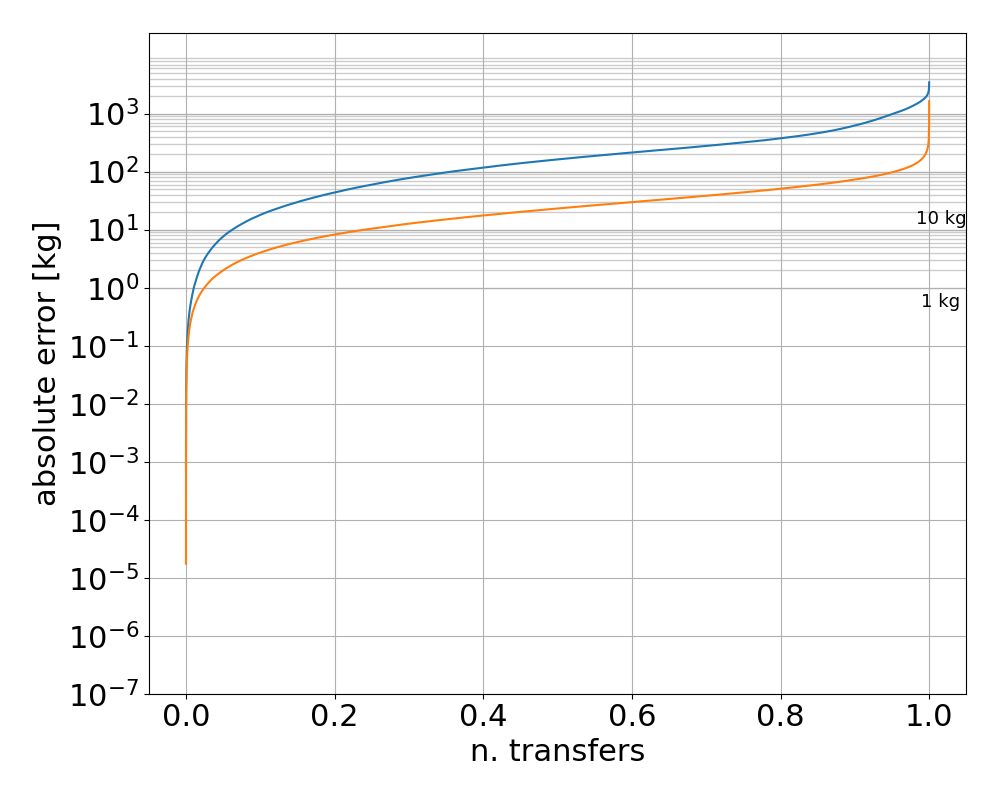}
    \caption{High time of flight (above 1500 days) database results.}
    \label{fig:abs_error_high_tof}
  \end{subfigure}
  \caption{Analytical and machine learning approximation results over $DB_1$, $DB_2$, $DB_3$, $DB_4$. The x-axis represents the number of transfers within a certain error threshold, normalized by the length of the database.}
  \label{fig:dbs_results_figures}
\end{figure*}

\begin{figure*}[hbt!]
  \centering
  \begin{subfigure}{0.49\textwidth}
    \includegraphics[width=\linewidth]{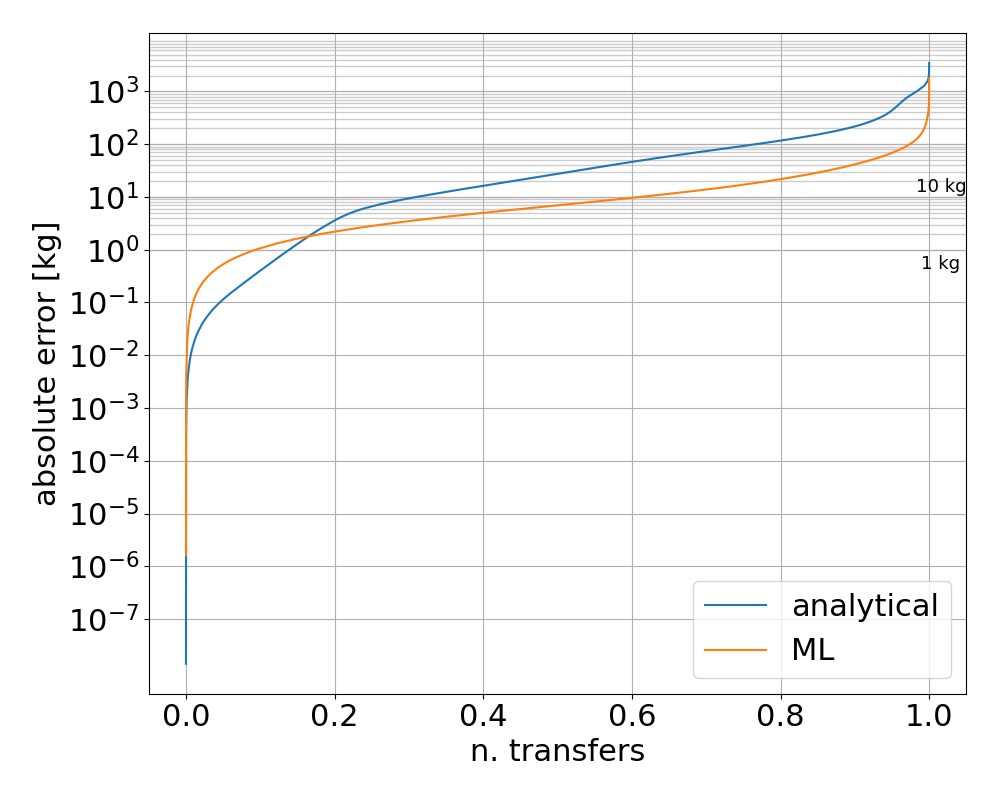}
    \caption{Sorted absolute error on the final mass estimation between analytical and ML approximation, over the whole database.}
    \label{fig:abs_error_all_tof}
  \end{subfigure}
  \hfill
  \begin{subfigure}{0.49\textwidth}
    \includegraphics[width=\linewidth]{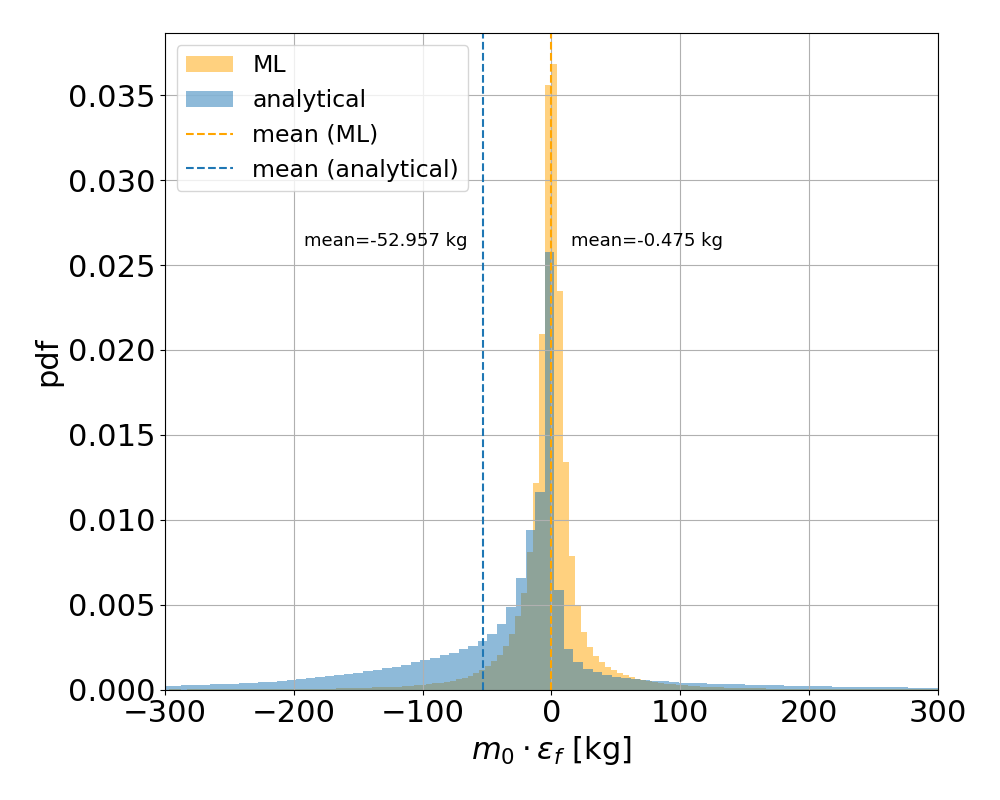}
    \caption{Relative error between the estimated and target final mass, times the initial mass.}
    \label{fig:hist_errors}
  \end{subfigure}
  \caption{Statistics of the errors of the analytical and ML approximations, over the whole database.}
\end{figure*}

In Fig.~\ref{fig:dbs_results_figures}, we display the absolute error in the final spacecraft mass, sorted from the lowest to the highest, for all four databases, starting with the low time of flight one in the upper left, the medium-low time of flight in the upper right, the medium-high time of flight in the lower left, and the high time of flight one on the lower right. The machine learning approximation consistently outperforms the analytical approximation. In particular, for the low, medium-low, and medium-high time of flights, the analytical approximation can achieve errors below the machine learning approximation for a small portion of the dataset size (roughly between 10 and 20\% of the cases). Instead, for the high time of flight cases, the machine learning approximation always has a better performance than the analytical one. 
In Fig.~\ref{fig:abs_error_all_tof}, we also display the same plot for the overall database, confirming that below absolute relative errors in the final mass of about 1.8 kg, the analytical approximation has an edge w.r.t. the machine learning one, but above that, the machine learning approximation achieves an enhanced accuracy, thereby confirming that in these scenarios, whenever a database of transfers is available, it is advisable to opt for a machine learning approach, as it is more capable of generalizing and maintaining lower errors. 
To have a better understanding of the error distribution of the machine learning and analytical approximation over the whole database, in Fig.~\ref{fig:hist_errors}, we display the probability density function of the initial mass of the transfer times the relative error (i.e., $\epsilon_f = (m_{f,target}-m_{f,estimated})/m_{f,target}$) for both cases. The x-axis is cropped between -300 and 300 kg, to improve the visual representation of the errors. As it is confirmed by these histograms, the neural network gives a more accurate and less biased estimation of the transfer cost. Moreover, as already mentioned in Sec.~\ref{sec:machine_learning_approximations}, the inference speed and data preparation (e.g. normalization of the inputs) also resulted in faster times w.r.t. the analytical approximation, with a speedup factor of roughly 50$\times$ (measured on a MacBook M1Pro). 

\section{Conclusions}

Having approximators that accurately and promptly estimate the low-thrust transfer cost among small bodies in the asteroid belt is crucial for the preliminary mission design of these tours. In particular, these techniques can help alleviate the cost of the optimization to find optimal sequences of asteroids to rendezvous with.

In this work, we have introduced a new analytical approximation building on the previously developed \textsc{MIMA} approximation~\cite{hennes2016fast}, by assuming that the low-thrust transfer can be approximated with two burn arcs and a middle coast arc (which goes to zero for time optimal transfers), whose lengths and thrust directions are to be solved for. Assuming constant thrust, an approximation is derived, which results in a good guess for the transfer cost estimation.
In parallel, we also computed the optimal control solution for a set of 3 million transfers in the asteroid belt, both time and fuel optimal, and we trained machine learning approximators to learn the transfer cost. The database is released open-source, and the comparison is both performed on different database splits, where the split is performed according to the time of flight, as well as over the entire dataset. It is shown that machine learning approaches can consistently outperform the analytical approximation, both in terms of accuracy and inference times (with speedups of about 50$\times$ with respect to the analytical approach), for all the cases. In particular, the analytical approximation remains very competitive, if not better, for the low time of flight cases and up to errors of about 1.8 kg in final mass. Nonetheless, for all the other scenarios, and especially for longer time of flights (e.g. multi-revolution transfers), the machine learning method has a significant edge, thereby confirming that it is convenient to use these techniques, whenever a database of transfers to learn from is available or can be computed.
\section{Data availability}
The database used in this work for training the machine learning approach, and for comparing the results and performing numerical experiments is available in Zenodo \url{https://zenodo.org/records/10972838}.

\bibliographystyle{ISSFD_v01}
\bibliography{references}

\begin{thebibliography}{10}
\newcommand{\enquote}[1]{``#1''}
\expandafter\ifx\csname urlstyle\endcsname\relax
  \providecommand{\doi}[1]{doi:\discretionary{}{}{}#1}\else
  \providecommand{\doi}{doi:\discretionary{}{}{}\begingroup \urlstyle{rm}\Url}\fi

\bibitem{demeo2015compositional}
DeMeo, F., Alexander, C., Walsh, K., Chapman, C., Binzel, R., et~al.
\newblock \enquote{The compositional structure of the asteroid belt.}
\newblock Asteroids iv, Vol.~1, p.~13, 2015.

\bibitem{minton2009record}
Minton, D.~A. and Malhotra, R.
\newblock \enquote{A record of planet migration in the main asteroid belt.}
\newblock Nature, Vol. 457, No. 7233, pp. 1109--1111, 2009.

\bibitem{bottke2015collisional}
Bottke, W.~F., Bro{\v{z}}, M., O’Brien, D.~P., Bagatin, A.~C., Morbidelli, A., and Marchi, S.
\newblock \enquote{The collisional evolution of the main asteroid belt.}
\newblock Asteroids IV, Vol.~1, pp. 701--724, 2015.

\bibitem{farinella1992collision}
Farinella, P. and Davis, D.~R.
\newblock \enquote{Collision rates and impact velocities in the main asteroid belt.}
\newblock Icarus, Vol.~97, No.~1, pp. 111--123, 1992.

\bibitem{russell2012dawn}
Russell, C. and Raymond, C.
\newblock \enquote{The dawn mission to Vesta and Ceres.}
\newblock The dawn mission to minor planets 4 vesta and 1 ceres, pp. 3--23, 2012.

\bibitem{chen2014accessibility}
Chen, Y., Baoyin, H., and Li, J.
\newblock \enquote{Accessibility of main-belt asteroids via gravity assists.}
\newblock Journal of Guidance, Control, and Dynamics, Vol.~37, No.~2, pp. 623--632, 2014.

\bibitem{grigoriev2013choosing}
Grigoriev, I. and Zapletin, M.
\newblock \enquote{Choosing promising sequences of asteroids.}
\newblock Automation \& Remote Control, Vol.~74, No.~8, 2013.

\bibitem{olympio2011optimal}
Olympio, J.~T.
\newblock \enquote{Optimal control problem for low-thrust multiple asteroid tour missions.}
\newblock Journal of guidance, control, and dynamics, Vol.~34, No.~6, pp. 1709--1720, 2011.

\bibitem{izzo2016designing}
Izzo, D., Hennes, D., Sim{\~o}es, L.~F., and M{\"a}rtens, M.
\newblock \enquote{Designing complex interplanetary trajectories for the global trajectory optimization competitions.}
\newblock Space Engineering: Modeling and Optimization with Case Studies, pp. 151--176, 2016.

\bibitem{vinko2008global}
Vink{\'o}, T. and Izzo, D.
\newblock \enquote{Global optimisation heuristics and test problems for preliminary spacecraft trajectory design.}
\newblock Advanced Concepts Team, ESATR ACT-TNT-MAD-GOHTPPSTD, 2008.

\bibitem{shen2023dyson}
Shen, H.-X., Luo, Y.-Z., Zhu, Y.-H., and Huang, A.-Y.
\newblock \enquote{Dyson sphere building: On the design of the GTOC11 problem and summary of the results.}
\newblock Acta Astronautica, Vol. 202, pp. 889--898, 2023.

\bibitem{edelbaum1961propulsion}
Edelbaum, T.~N.
\newblock \enquote{Propulsion requirements for controllable satellites.}
\newblock Ars Journal, Vol.~31, No.~8, pp. 1079--1089, 1961.

\bibitem{casalino2007improved}
Casalino, L. and Colasurdo, G.
\newblock \enquote{Improved Edelbaum's approach to optimize low earth/geostationary orbits low-thrust transfers.}
\newblock Journal of guidance, control, and dynamics, Vol.~30, No.~5, pp. 1504--1511, 2007.

\bibitem{li2020deep}
Li, H., Chen, S., Izzo, D., and Baoyin, H.
\newblock \enquote{Deep networks as approximators of optimal low-thrust and multi-impulse cost in multitarget missions.}
\newblock Acta Astronautica, Vol. 166, pp. 469--481, 2020.

\bibitem{pontryagin1962maximum}
Pontryagin, L., Boltyanskii, V., Gamkrelidze, R., and Mishchenko, E.
\newblock \enquote{The maximum principle.}
\newblock The Mathematical Theory of Optimal Processes. New York: John Wiley and Sons, 1962.

\bibitem{sims1997preliminary}
Sims, J.~A. and Flanagan, S.~N.
\newblock \enquote{Preliminary design of low-thrust interplanetary missions.}
\newblock 1997.

\bibitem{gill2005snopt}
Gill, P.~E., Murray, W., and Saunders, M.~A.
\newblock \enquote{SNOPT: An SQP algorithm for large-scale constrained optimization.}
\newblock SIAM review, Vol.~47, No.~1, pp. 99--131, 2005.

\bibitem{biscani2020parallel}
Biscani, F. and Izzo, D.
\newblock \enquote{A parallel global multiobjective framework for optimization: pagmo.}
\newblock Journal of Open Source Software, Vol.~5, No.~53, p. 2338, 2020.

\bibitem{mereta2017machine}
Mereta, A., Izzo, D., and Wittig, A.
\newblock \enquote{Machine learning of optimal low-thrust transfers between near-earth objects.}
\newblock \enquote{International Conference on Hybrid Artificial Intelligence Systems,} pp. 543--553. Springer, 2017.

\bibitem{izzo2019machine}
Izzo, D., Sprague, C.~I., and Tailor, D.~V.
\newblock \enquote{Machine learning and evolutionary techniques in interplanetary trajectory design.}
\newblock Modeling and Optimization in Space Engineering: State of the Art and New Challenges, pp. 191--210, 2019.

\bibitem{viavattene2022artificial}
Viavattene, G. and Ceriotti, M.
\newblock \enquote{Artificial neural networks for multiple NEA rendezvous missions with continuous thrust.}
\newblock Journal of Spacecraft and Rockets, Vol.~59, No.~2, pp. 574--586, 2022.

\bibitem{hennes2016fast}
Hennes, D., Izzo, D., and Landau, D.
\newblock \enquote{Fast approximators for optimal low-thrust hops between main belt asteroids.}
\newblock \enquote{2016 IEEE Symposium Series on Computational Intelligence (SSCI),} pp. 1--7. IEEE, 2016.

\bibitem{acciarini_2024_10972838}
Acciarini, G., Izzo, D., and Beauregard, L.
\newblock \enquote{{Optimal low thrust transfers among asteroid belt asteroids.}}, Apr. 2024.
\newblock \doi{10.5281/zenodo.10972838}.

\bibitem{izzo_gtoc12_2024}
Izzo, D., Märtens, M., Beauregard, L., Bannach, M., Acciarini, G., Blazquez, E., Hadjiivanov, A., Grover, J., Hei{\ss}el, G., Shimane, Y., and Yam, C.~H.
\newblock \enquote{Asteroid Mining: ACT\&Friends' Results for the GTOC~12 Problem.}
\newblock \textit{to appear in} Astrodynamics, 2024.

\end{thebibliography}

\end{document}